# The Role of Spin-Flip Collisions in a Dark Exciton Condensate

Subhradeep Misra[1], Michael Stern[2], Vladimir Umansky[1] and Israel Bar-Joseph[1]

[1]*Department of Condensed Matter Physics, Weizmann Institute of Science, Rehovot, Israel*

[2]*Department of Physics and Center for Quantum Entanglement Science and Technology, Bar-Ilan University, Ramat-Gan, Israel*

**Abstract**

We show that a Bose-Einstein condensate consisting of dark excitons forms in GaAs coupled quantum wells at low temperatures. We find that the condensate extends over hundreds of micrometers, well beyond the optical excitation region, and is limited only by the boundaries of the mesa. We show that the condensate density is determined by spin flipping collisions among the excitons, which convert dark excitons into bright ones. The suppression of this process at low temperature yields a density buildup, manifested as a temperature-dependent blueshift of the exciton emission line. Measurements under in-plane magnetic field allows us to preferentially modify the bright excitons density, and determine their role in the system dynamics. We find that their interaction with the condensate leads to its depletion. We present a simple rate equations model, which well-reproduces the observed temperature, power, and magnetic field dependence of the exciton density.

In the dilute limit, excitons can be considered as a Bose gas, and are thus expected to undergo a Bose Einstein condensation (BEC) transition at low temperatures [1, 2]. Over the past few decades, there have been many attempts to realize such a condensate in various material systems. A particularly interesting research direction, which attracted considerable attention, is using indirect excitons (IX) in coupled quantum wells (CQW) structures [3 - 6]. An attractive feature of this system is its long lifetime, which allows achieving the critical condensation density at moderate excitation powers, and establishing quasi-equilibrium at low temperatures. Indeed, a bulk of experimental works on CQW reported a behavior that is consistent with an exciton BEC transition at low temperatures [7-17].

An important development in the quest for the exciton BEC was introduced a little over a decade ago by Combescot *et al*. [18, 19], who argued that the condensate should be dark. Their argument was based on the fact that electron-hole exchange interaction removes the degeneracy between the bright ($J_z = \pm 1$) and dark ($J_z = \pm 2$) exciton energies, such that the dark exciton is lower in energy than the bright one, and hence, it should be the ground state for condensation. Indeed, a series of studies performed in the last decade provided evidence for the formation of a dark condensate [9, 11, 12, 14, 16], with the most compelling one coming from measurement of the blueshift of the IX peak energy as the temperature is decreased at constant illumination power [12]. This blueshift, $\Delta E$, is due to the dipole-dipole repulsive interaction among excitons and is proportional to their density [20]. Hence, its increase at low temperatures can be interpreted as a signature of a buildup of long-lived dark excitons at the ground state of the system.

The unique composition of the system, of a thermal part consisting of short-lived bright excitons, and a condensate of long-lived dark excitons, poses a challenge in understanding its formation dynamics. It can be viewed as an intermediate case between an atomic BEC, with a fixed number of particles, and a cavity-polaritons condensate, whose steady-state density is set by its fast radiative recombination. Indeed, these conflicting views are manifested in the literature, where some consider it as an equilibrium system [5, 6], while others argue that its density is determined by nonradiative recombination processes [21].

In this work, we address this challenge by conducting spectroscopic studies of GaAs/AlGaAs CQW in a mesa structure. We first present conclusive evidence that a dark condensate is formed in this structure at low temperatures, and extends over hundreds of micrometers (µm), well beyond the optical excitation region. We then study its formation dynamics and show that its steady-state density is determined by spin flipping collisions between excitons, which convert dark excitons into bright ones. We show that the inhibition of these processes at low temperatures, $T < 3K$, gives rise to the condensate density buildup.

**Evidence for dark exciton condensation**

The sample structure is illustrated in Fig. 1. It consists of a coupled quantum well (CQW) system with two GaAs wells having widths of 12 and 18 nm, respectively, separated by an Al$_{0.28}$Ga$_{0.72}$As barrier of thickness 3 nm. The CQW is embedded in a 2 μm thick $n^+ - i - n^+$ structure having top and bottom silicon doped n$^+$ layers of Al$_{0.12}$Ga$_{0.88}$As ($n \sim 10^{18}$ cm$^{-3}$), that allows application of a voltage perpendicular to the growth direction. The different well width allows us to conduct spectroscopic measurements in each well. The sample was processed using optical lithography, isotropic wet etching and metal deposition technique to form rectangular mesas of dimension $200 \times 100\ \mu m^2$. Tunneling of the photoexcited carriers gives rise to charge separation and formation of a dipolar exciton. An exemplary spectrum, depicting the PL emission from the sample is shown in Fig. 1.

We begin by presenting several measurements that provide compelling evidence for the accumulation of dark exciton at low temperatures in our sample. Figure 2a shows the IX peak energy, $E_{IX}$, as a function of temperature at three gate voltages, $V_g$, while keeping the pump power fixed at $P = 10$μW. It is seen that $E_{IX}$ shifts to high energies with decreasing temperatures, reaching $\Delta E = 10 - 20$ meV at $T < 1$K. These large values of $\Delta E$ imply that the steady state exciton density, $n$, increases considerably as the temperature is lowered under constant excitation power, approaching $10^{11}$ cm$^{-2}$ at the lowest temperatures.

To confirm the presence of a high IX density, we measure the absorption spectra of the narrow and wide wells. This is done by tuning the energy of the excitation laser near the direct exciton resonance of each well, and measuring the generated photocurrent as a function of laser energy. Figure 2b shows the resulting narrow well spectrum as a function of temperature at -3.3V. The shift of the direct exciton line to higher energies with decreasing temperatures and the broadening of the lower energy trion line are clear manifestations of the presence of high density of electrons in this well. Measurements conducted in the wide-well are presented in section VII of SI appendix, and show a similar temperature dependence, confirming that a high hole density builds up in the wide well too. Solving the self-consistent Schrodinger-Poisson equation, assuming equal charges in the two wells, we could reproduce the shift of the direct exciton and obtain a density of ~$5 \times 10^{10}$ cm$^{-2}$ at the lowest temperature, in good agreement with $\Delta E = 10$meV obtained at Fig. 2a for this voltage.

To determine whether this increase of density is due to accumulation of bright or dark excitons we measure the corresponding change of the radiative rate, $\gamma_r$. By conducting measurements at various excitation powers and temperatures we find that $\gamma_r$ depends primarily on $E_{IX}$, and becomes faster as $E_{IX}$ increases (Fig. 2c), reflecting the increased overlap of the electron and hole wavefunctions as the external electric field is screened. We can therefore conclude that $\gamma_r$ increases as the temperature is lowered (and $E_{IX}$ increases), and hence, the steady-state bright exciton density should *decrease* with decreasing temperature. Furthermore, if we estimate the steady-state density that can be generated at 10 $\mu$W pump power using the measured lifetimes, which vary between $20 - 100$ ns over the relevant $\Delta E$ range, we obtain $0.4 - 2 \times 10^9$ cm$^{-2}$, far less than the density required to generate $\Delta E$ of $10 - 20$ meV. It follows that bright excitons buildup cannot explain the observed blueshift, suggesting that excitons, formed by electrons in the narrow well and holes in the wide well, with a much longer lifetime, accumulate at low temperatures. We confirmed that the electrons and holes are bound in an excitonic state by measuring the behavior of the IX line in a perpendicular magnetic field: A clear quadratic diamagnetic shift is observed (Fig. S4 of the SI appendix).

A conclusive proof for the accumulation of dark excitons is provided by measuring the blueshift away from the excitation beam. Figure 2d shows the normalized $\Delta E$ of the IX peak energy as a function of position along a horizontal cross-section of the mesa, $\Delta E(x)/\Delta E(x=0)$, where $x=0$ is the center of the excitation beam. The measured PL lines as a function of position are given in Fig. S9 of the SI. Since $\Delta E$ is proportional to the exciton density, this measurement yields the density distribution. It is seen that at high temperatures $\Delta E(x)$ follows the illumination beam profile, which is marked by the solid red line (the small increase in the distribution width is due to diffusion). Remarkably, however, as the temperature is lowered below ~6K, $\Delta E(x)$ becomes *constant* across the 200μm wide mesa, while the PL intensity, $I(x)$, continues to follow approximately the illumination beam profile. In fact, we find that the constant $\Delta E$ extends a few hundreds μm further into the leads to the mesa. The lack of correlation between $I(x)$ and $\Delta E(x)$ implies that long-lived particles, which contribute to $\Delta E$ but do not emit light, diffuse uniformly throughout the large mesa.

The exchange splitting, $\varepsilon_{bd}$, which is the energy difference between bright and dark excitons, is much smaller than $k_B T$ in GaAs CQW throughout the temperature range studied in this work. Hence, the formation of a large steady-state density of dark excitons implies that a high occupation of the ground state occurs at low temperatures, an indication for a BEC transition. This is corroborated by observation of residual spatial coherence over short lengths in an experiment performed in a trap geometry [11].

**The condensate loss mechanisms**

The question we wish to focus on in the rest of this paper is what determines the density of the condensate. In Fig. 3a we present an Arrhenius plot of the data of Fig. 2a, showing $\log(\Delta E_c)$ versus $1/T$. In order to consider the contribution of the dark condensate only, $\Delta E_c$, we subtract $E_{IX}(P,\infty)$, the IX energy at the same power and high temperatures ($T = 9$K), from the measured $E_{IX}(P,T)$, such that $\Delta E_c = E_{IX}(P,T) - E_{IX}(P,\infty)$. This removes the bright exciton contribution, which should be present also at high temperatures. A clear linear behavior is seen for the three voltages with an identical slope. This is a signature of an activated behavior, $\Delta E_c = A\exp(U_0/k_B T)$, and the identical slopes imply that the activation energy is the same at all voltages. (Clearly, the prefactor $A$ is the blueshift at $T = \infty$). Repeating this procedure for the data collected at other excitation powers reveals a similar behavior with the same activation energy (inset of Fig. 3a). This behavior indicates the existence of a potential barrier, such that when the dark excitons can be thermally excited above it – they are lost and $\Delta E$ decreases. Since we do not observe a corresponding drop in the PL intensity as the temperature is reduced, we conclude that the dark excitons are eventually converted into bright ones, and are quickly removed from the system by radiative recombination. This suggests that spin flipping collisions, which convert two dark excitons, $|\pm 2\rangle$, into two bright ones, $|\pm 1\rangle$, through particle exchange, $|+2\rangle + |-2\rangle \rightarrow |+1\rangle + |-1\rangle$, is the loss mechanism determining the condensate density, $n_c$.

We can express the rate of this particle exchange collision process as $\alpha n_c^2$, where $\alpha = \alpha_0 \exp(-U_0/k_B T)$, and $\alpha_0$ is the exchange energy density. To extract $U_0$ from the slope of the linear fits in Fig. 3a, we need to express $\Delta E$ in terms of $n_c$. We note that for a condensate $\Delta E \propto n_c^{3/2}$ [20], and hence $U_0$ is given by 3/4th of that slope. The resulting value for $U_0 \approx 0.5$meV corresponds to the height of the screened exciton-exciton repulsive potential at $\sim 2a_B$, where $a_B$ is the exciton radius [22]. When $k_B T > U_0$, excitons can be excited above this barrier and tunnel into the region $r < 2a_B$, and the probability for particle exchange substantially increases.

A confirmation to this picture is provided by examining the power dependence of the blueshift, $\Delta E(P)$, shown in Fig. 3b. It is seen that $\Delta E$ is sublinear in $P$ at all temperatures. Since $\Delta E$ should grow with $n_c$, it implies that the loss mechanism should be superlinear in $n_c$. Indeed, if we write the rate equation for $n_c$ as $dn_c/dt \approx P - \alpha n_c^2$, we recover the sublinear dependence of $\Delta E$ on $P$. We can use the resulting expression for the steady state condensate density, $n_c \approx \sqrt{P/\alpha}$, to obtain the exchange energy density, $\alpha_0$. We find $\alpha_0 \sim 10^{-8}$ m²s⁻¹, increasing by a factor of 3 between $V_g = -2.5$ to $-3.7$V. This order of magnitude is in a very good agreement with the calculation of Ref. [21] for the exchange integral using $d = 18$nm, the dipole length in our sample, thereby providing a strong support for this interpretation. We note that the characteristic lifetime for this process is $\sim \sqrt{P\alpha}$, and is a few µs for $P = 10$µW.

Rewriting $\alpha = \alpha_0 \exp(-\frac{T_0}{T})$ we can define a characteristic temperature, $T_0 \approx 7$K, below which this scattering process is exponentially suppressed and the dark exciton density increases. Hence, the increase in $\Delta E$ with reducing temperatures should not be interpreted as the onset of BEC, but is rather a manifestation of the exponential inhibition of the two-particle scattering process at $T \ll T_0$. In other words, the large density of dark excitons, which is observed at low temperatures, implies that a macroscopic occupation of the ground state does occur in this system. However, its diminishing at higher temperatures is not related to exceeding the critical temperature for condensation. This can explain why the temperature below which the blueshift starts to increase is rather constant, and does not scale linearly with power as expected for the critical temperature for BEC.

Earlier works that have studied the impact of particle exchange on the condensate [19, 21], focused on the critical density beyond which the condensate becomes 'gray", e.g – acquires a bright component. At this critical density the energy correction due to exchange processes between excitons becomes larger than $\varepsilon_{bd}$, and the BEC is a coherent mixture of dark and bright components. While this is true for $T = 0$, at a finite temperature these particle exchange processes should be active also at densities lower than that critical density, and become the dominant loss mechanism of the condensate.

Finally, we wish to emphasize that in most of the power and temperature range, the role of nonradiative recombination processes is insignificant in determining the condensate density. Estimate of the nonradiative recombination rate in our sample, $\gamma_{nr} \approx 3 \times 10^3$ s$^{-1}$, is given in section VII of the SI appendix, and it is easy to see that $\alpha n_c^2 \gg \gamma_{nr} n_c$ for $n_c \gtrsim 10^{10}$cm$^{-2}$ and $T \gtrsim 1$K. Hence, only at very low powers and temperatures, when spin flipping processes are suppressed, nonradiative processes become important [21].

**The condensate - bright excitons interaction**

To uncover the role of the $J_z = \pm 1$ bright excitons, we preferentially modify their density, $n_1$, and study the response of the system to this modification. This is achieved by applying a magnetic field parallel to the CQW layers, $\boldsymbol{B}_{\parallel} = \hat{\boldsymbol{x}} B$. Under such field, the exciton dispersion in the direction normal to it, $\hat{\boldsymbol{y}}$, is shifted, such that its minimum is at $k_y = eBd/\hbar$. At $B > 1$T this minimum resides outside the radiative zone, e.g - $eBd/\hbar > k_0 = E_{IX}\sqrt{\varepsilon}/\hbar c$, and only a fraction of the excitons can recombine radiatively. Hence, the effective radiative lifetime of the system, $\tau_r^{eff}$, and consequently $n_1$, are expected to grow with $B$ [22]. Indeed, we find a 40 fold increase in the measured lifetime, from 40ns at $B = 0$ to $\approx 1.6\mu$s at $B = 4$T (Fig. S2 of the SI appendix). To see the effect of this increment in $n_1$ we examine the behavior at high temperatures, $T > 2K$, where the contribution of the dark excitons to the blueshift is relatively small. Figure 4a shows $\Delta E(B)$ at $T = 2.9$K for two voltages, $V_g = -2.5$V and $-2.9$V, where we have subtracted the zero-field blueshift due to dark excitons. Indeed, a clear blueshift of the IX energy is seen as the magnetic field is ramped between 0 and 4T. As expected, the longer zero-field lifetime at $V_g = -2.9$V gives rise to a larger blueshift at 4T. This magnetic field-induced $\Delta E$ represents a straightforward steady-state exciton accumulation, $n_1 \sim P \tau_r^{eff}$, as opposed to the dark exciton accumulation at $B = 0$, which is not lifetime related.

A surprising effect is observed at lower temperatures, where $n_c$ becomes significant: We find that $\Delta E(B)$ *decreases* at low fields, reaches a minimum at $B \sim 2.5$T, and then rises again at higher fields, such that the IX energy at $B = 4$T is the same for all temperatures (Fig. 4b). Dark – bright exciton mixing could be ruled out as the cause of this dip: The effect of such mixing should be observed already at $B_{\parallel} \ll 1$T, however, as can be seen in Fig. 4b, we don't find any significant drop of the IX energy below 1T (See expanded discussion in section III of the SI appendix).

This behavior, which is verified to appear in other voltages and powers, can be interpreted as a competition between two opposing trends: An increase of $\Delta E$ due to the buildup of the $|\pm 1\rangle$ bright exciton density, $n_1$, and a corresponding decrease due to the depletion of the $|\pm 2\rangle$ dark exciton density, $n_c$, with increasing magnetic field. We note that the magnitude of the dip in $\Delta E(B)$, which is not present in Faraday configuration measurements, becomes larger with decreasing temperature, as the zero-field dark exciton density, $n_c(B=0)$, increases. (The data for Faraday configuration is presented in section V of the SI appendix). We can therefore conclude that the depletion of the condensate is proportional to both $n_1$ and $n_c$, and can be formally expressed as $\beta n_1 n_c$.

Such a loss term represents interaction between bright and dark excitons. Clearly, particle exchange does not yield a net change in the dark exciton density in this case. The underlying mechanism can be understood in the following way: It is well known that the dominant spin flip mechanism of excitons in CQW, is the Dyakonov-Perel (DP) effect [24]. In the single particle picture of this effect, momentum scattering with impurities changes the direction of the exciton motion in the non-centrosymmetric crystal, and gives rise to loss of spin orientation. We suggest that momentum scattering of the dark excitons by the thermal bright excitons, with a rate that can be expressed as $\beta n_1 n_c$, gives rise to a similar effect. It is easy to see that inclusion of this term in the zero-field behavior would yield a saturation of $n_c$ with power at low temperatures, where the $\alpha n_c^2$ term is suppressed: If we express $n_1$ as $\approx P/\gamma_r$ and write the rate equation as $dn_c/dt \approx P - \beta n_1 n_c$, we can see that at steady state, $n_c \approx \frac{\gamma_r}{\beta}$. This explains the origin of the saturation of the $\Delta E(P)$ curve at high power, which is observed at $T < 1.5K$ (Fig. 3b).

## Discussion

The findings presented above allow us to construct a nonlinear rate equations model (See section I of the SI appendix), which includes the $\alpha n_c^2$ and $\beta n_1 n_c$ terms, for the full system, comprising of thermal bright and dark excitons, with steady state densities $n_1$ and $n_2$, respectively, and a condensate of dark excitons, with a steady state density $n_c$. The fast decay of the bright excitons through radiative recombination acts as a sink for the system, thereby limiting $n_c$. The particles exchange between the three reservoirs, $n_1$, $n_2$ and $n_c$, and with the environment, through radiative and nonradiative recombination, $\gamma_r$ and $\gamma_{nr}$, respectively, is schematically depicted in Fig. 5a. The thermal bath is pumped at a rate $P$, and the condensate is formed through stimulated scattering at a rate $W^+ = R n_2 n_c$. Solving these equations we obtain the steady-state density of $n_1$, $n_2$, and $n_c$ as a function of power and temperature (See section II and III of the SI appendix). The total density $n(P,T) = n_1 + n_2 + n_c$ is depicted in Fig. 5b, where the increase of $n$ with decreasing temperature at constant power, and the sublinear dependence of $n$ on $P$, are clearly seen. The solid lines in Figs. 3b and 4b are best fit to $\Delta E$ predicted by this model. It is seen that the fitted lines are in very good agreement with the experimental data. The dark nature of the condensate is therefore understood as a result of the unique dynamical properties of the system: the long lifetime of the dark exciton and the suppression of dark-bright spin flip processes.

In the limit of low power and temperatures, where dark to bright spin flipping processes are suppressed, the condensate lifetime is determined by nonradiative recombination, and our model converges with that suggested in [21]. Indeed, the steep rise of the blueshift with power at $T \leq 1K$ seen in Fig. 5b is a manifestation of this effect.

Similar nonlinear loss mechanisms due to interactions among the condensate particles and between them and the thermal bath, are known to play important role in atomic and cavity-polariton condensates [25, 26]. In that sense, and in its non-equilibrium nature, the system of dipolar exciton in CQW is very similar to the system of cavity-polaritons. We note, however, that the interactions are much stronger in CQW, due to the large electric dipole associated with the IX, resulting in strong correlation effects [6]. Unfortunately, its dark nature makes it difficult to study using spectroscopic tools.

## Materials and Methods

The experiments are conducted in a dilution refrigerator with optical windows ($0.05 < T < 6$ K), a variable temperature pumped helium cryostat ($1.5$ K $< T$) and a variable temperature 8T split-coil magneto-optic cryostat ($1.5\ K < T$). This allows us to cover the temperature and magnetic field range of interest of this work, $0.1 < T < 10$ K and $0 < B < 4$ T. For most of the experiments, a Gaussian beam with a 60 μm radius (half width at half maximum) of a continuous wave laser at an energy 1.590 eV illuminates the sample. This energy is well below the barrier height, ensuring carriers are excited in the wells only, and carrier depletion effects are avoided [27]. The dark current through the sample at the relevant voltage range is ~100 pA, and the photocurrent at the highest excitation power, 100 μW, is below 1 nA. The photoluminescence (PL) from the sample is guided to a U-1000 Jobin-Yvon Raman spectrometer to collect the spectra and analyze it, and is imaged on a cooled EMCCD camera (Andor iXon Ultra) with a spatial resolution of 3μm. Spatially resolved spectroscopic measurements are performed using a pinhole having diameter 20 μm placed at the focal plane of the collection objective. This allows selective collection of emitted light from various distinct areas of the mesa for imaging and spectroscopic analyses.

We carry out time-resolved measurements to determine the lifetime of indirect excitons using a correlated photon counting system. To pulse the CW laser we use an acousto-optic modulator (AOM) with a rise time of 17 ns, and an Avalanche Photo-Diode (APD) is used to detect the photon events. The pulse duration is 1μs, in a 10μs duty cycle. The pulses to the AOM and the measurement time windows at the APD are synchronized by a fast arbitrary waveform generator (Tabor AWG WX1284) and a National Instruments Data Acquisition card (NIDAQ) is used for counting the APD output.

## Acknowledgments

This work is supported by the Israeli Science Foundation, Grant No. 2139/20, and by the Minerva foundation.

# Figures

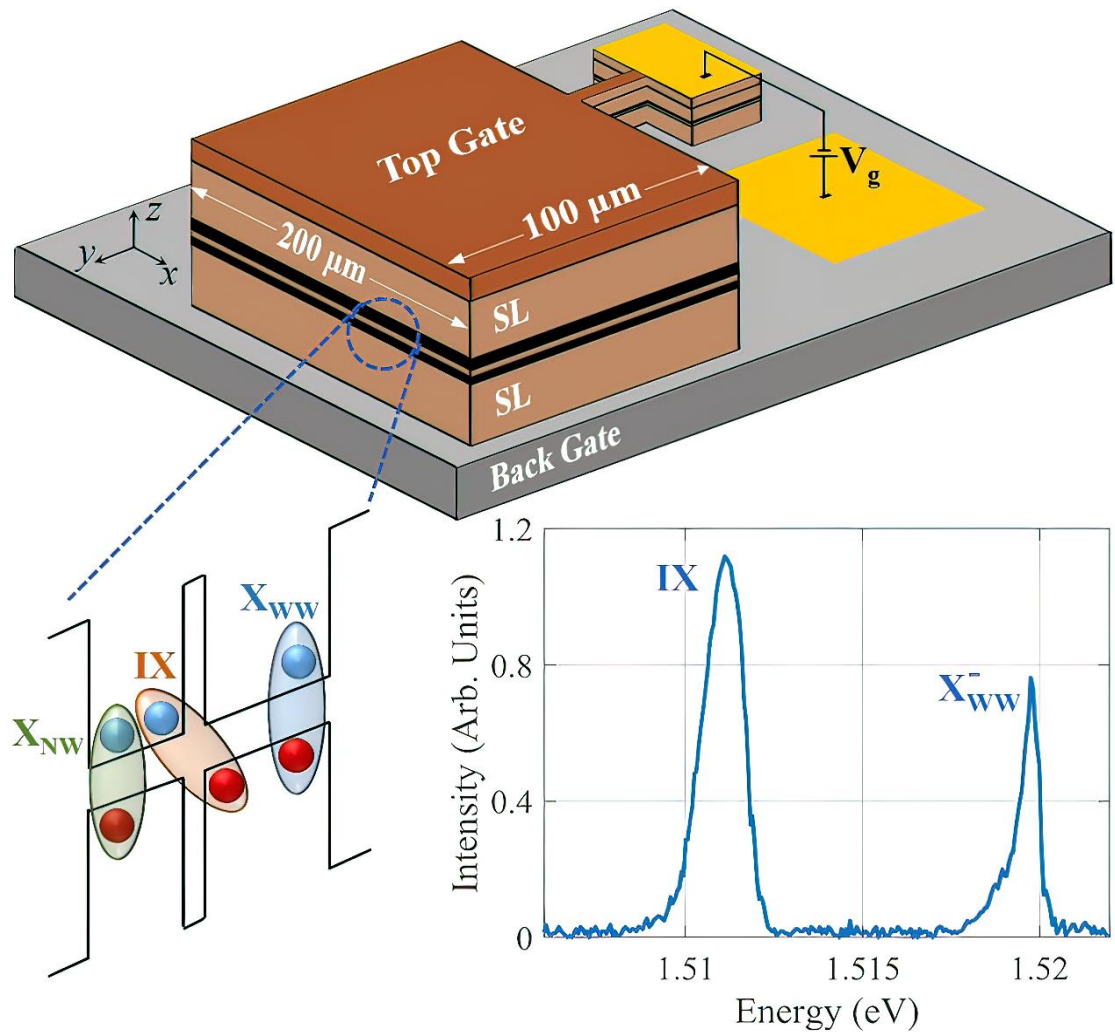

**Figure 1**: Upper panel: Schematic diagram of the sample structure. The black stripes at the middle region denote the CQW. Lower left planel shows the band diagram of this region under a negative gate voltage. Lower right panel is an exemplary PL spectrum, taken at $V_g = -3.3$ V, P = 10 µW, and T = 1.7 K. The IX and the negatively charged wide well trion ($X_{WW}^-$) are labelled.

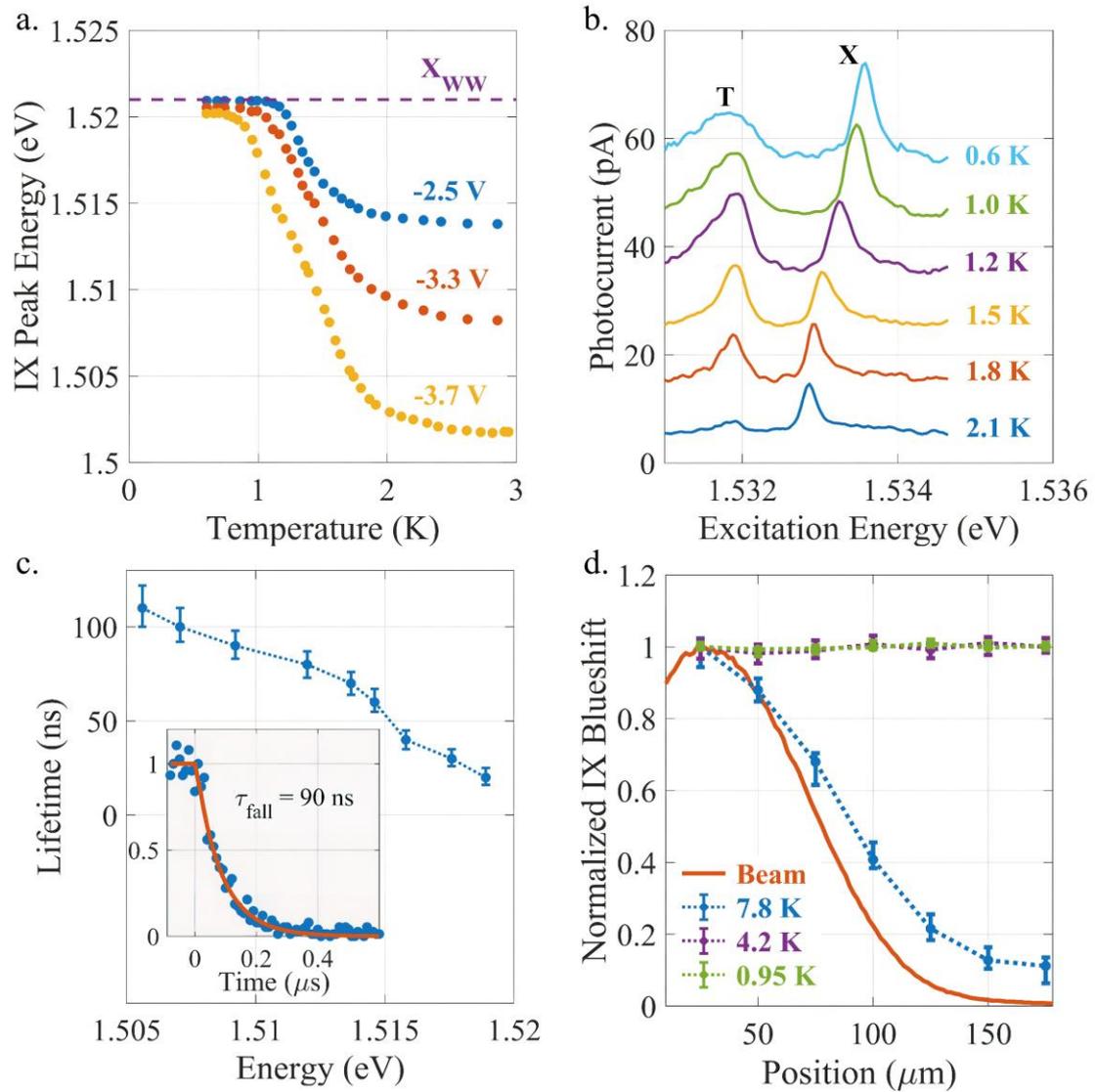

**Figure 2**: a. The IX peak energy as a function of temperature at three gate voltages and a fix power P = 10μW. The dashed line marks the energy of the wide-well exciton $X_{WW}$; b. Photocurrent spectra as the illuminating laser energy is tuned around the narrow-well exciton resonance at various temperatures. Here $V_g = -3.3$ V and P = 2 μW; c. The IX radiative lifetime as a function of peak energy, $E_{IX}$. (inset) An exemplary trace of the lifetime measurement (blue data points) using correlated photon counting. The red line is a fit to exponential decay, with a fall time of 90 ns; d. The IX blueshift along a horizontal cross-section of the mesa at three temperatures, normalized by its value at the center of the excitation beam. (The excitation beam profile is shown by the red solid line.) The transition to a constant blueshift at low temperatures can be clearly observed.

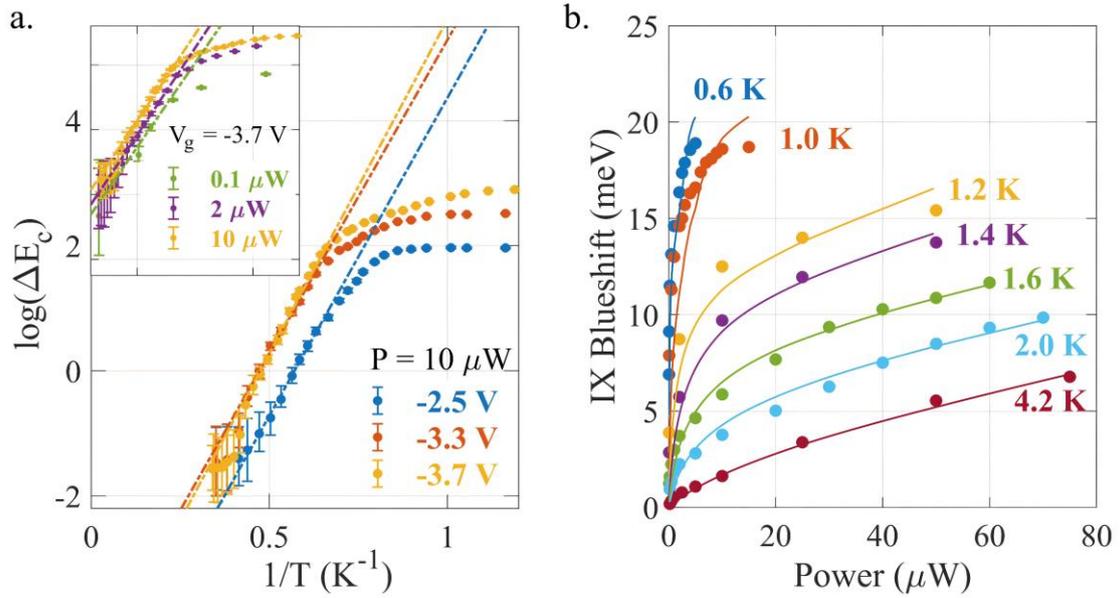

**Figure 3**: a. Arrhenius plot showing $\log(\Delta E_c)$ versus $1/T$, where $\Delta E_c = E_{IX}(P, T) - E_{IX}(P, \infty)$ is the blueshift due to the condensate only at $P = 10\mu W$. $E_{IX}(P, T)$ is taken from the data in Fig. 2a, and $E_{IX}(P, \infty)$ is the IX energy at the same power and high temperature ($T = 9K$). The dash-dotted lines are linear fits for the data points in the region before the crossing with $X_{WW}$. Remarkably, the fits give practically same slopes. (inset) The same analysis at $V_g = -3.7V$ for various powers (the x and y scales are the same as in the main figure); b. Filled circles - the measured power dependence of the IX blueshift $\Delta E$ at various temperatures at $V_g = -3.7V$. The maximal $\Delta E = 18$meV is limited by the crossing of the $X_{WW}$ line (See Fig. 2a). Solid lines - the steady-state solutions of the rate equations model.

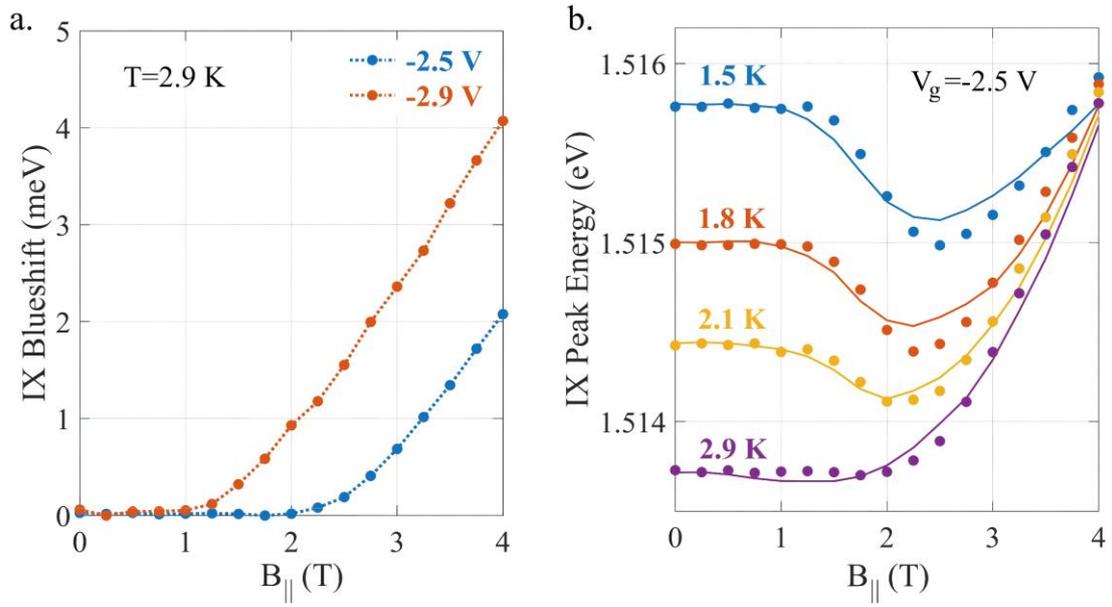

**Figure 4**: a. Measured ΔE as a function of in-plane magnetic field at two voltages at P = 10μW; b. Dots - measured ΔE as a function of $B_{||}$ at several temperatures. Solid lines - the calculated behavior using the rate equations model.

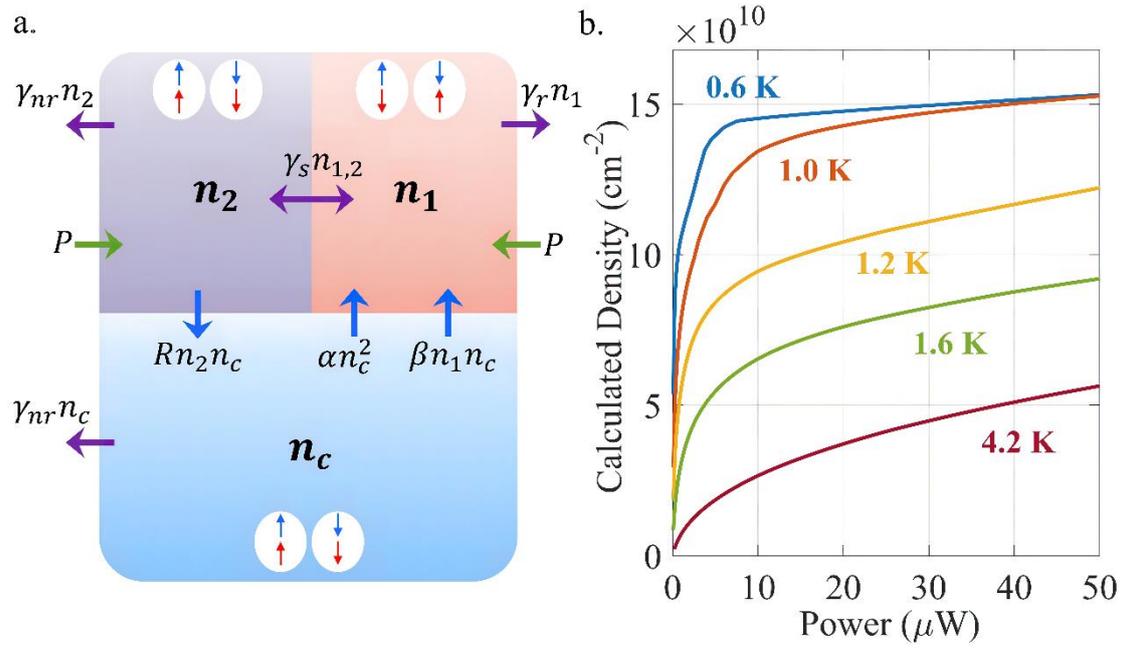

**Figure 5**: a. Schematic representation of the particles exchange among the three reservoirs, $n_1$, $n_2$, and $n_c$, and with the environment. The spin orientations of the electron and hole within the excitons are marked by the blue and red arrows, respectively. b. The calculated steady-state total exciton density as a function of excitation power at a few temperatures.

# The Role of Spin-Flip Collisions in a Dark Exciton Condensate

## Supplementary Information

Subhradeep Misra[1], Michael Stern[2], Vladimir Umansky[1] and Israel Bar-Joseph[1]

[1]*Department of Condensed Matter Physics, Weizmann Institute of Science, Rehovot, Israel*
[2]*Department of Physics and Center for Quantum Entanglement Science and Technology, Bar-Ilan University, Ramat-Gan, Israel*

# Table of Contents

## I. The Rate Equations Model

Here we describe the rate equations model used to explain the temperature and magnetic field dependent IX energy shift. As mentioned in the text, at steady state we can classify the excitons in the system into two groups based on spin: $J_z = \pm 1$ that can recombine radiatively and $J_z = \pm 2$ that cannot couple to light due to angular momentum conservation law and may only recombine through slow non-radiative processes. Since only $J_z = \pm 2$ states can condense, we assume these excitons to be either thermal or condensed, whereas the $J_z = \pm 1$ excitons are thermal only. Hence, we have three reservoirs in the system: thermal $J_z = \pm 1$ excitons (with density $n_1$), thermal long-lived $J_z = \pm 2$ excitons ($n_2$) and condensed $J_z = \pm 2$ excitons ($n_c$). The particles exchange among the three reservoirs and with the environment are schematically depicted in Fig. 5a of the main text and discussed there. These considerations lead to the following system of equations:

$$\frac{\partial n_1}{\partial t} = P - \gamma_r n_1 - \gamma_s n_1 + \gamma_s n_2 + \alpha n_c^2 + \beta(n_1 + n_2)n_c \qquad (1)$$

$$\frac{\partial n_2}{\partial t} = P - R n_2 n_c - \gamma_{nr} n_2 - \gamma_s n_2 + \gamma_s n_1 \qquad (2)$$

$$\frac{\partial n_c}{\partial t} = R n_2 n_c - \alpha n_c^2 - \beta(n_1 + n_2)n_c - \gamma_{nr} n_c \qquad (3)$$

Here $P$ is the pumping power, $\gamma_r$ and $\gamma_{nr}$ are the radiative and non-radiative decay rate of the excitons, and $\gamma_s$ is the spontaneous spin flip rate between $n_1$ and $n_2$. $R$ is the stimulated scattering rate from $n_2$ to $n_c$ [1], $\alpha$ is the two-particle scattering rate, and $\beta$ is the scattering rate of $n_c$ to $n_1$ due to the rapid recombination of the dipolar bright excitons.

An important assumption here is that a condensate exciton cannot flip its spin spontaneously (at a rate $\gamma_s n_c$), but can only do that through scattering with another condensate exciton (at a rate $\alpha n_c^2$) or with a thermal exciton at a rate $[\beta(n_1 + n_2)n_c]$. Note that in the equations above we do not consider the spatial dependence of the various densities. Clearly, the spatial dependence of $n_1$ and $n_c$ is very different: While $n_1$ should follow approximately the excitation beam profile, $n_c$ is found to be constant throughout the mesa. We have repeated the measurements using a broad excitation beam, which is almost uniform across the mesa, and obtained very similar results. In fact, we find that the measured blueshift depends on the total power and is insensitive to the excitation beam profile. We can therefore conclude that this difference between the spatial profiles does not play an important role.

## II. Approximated Solutions

### a. The Condensate at Zero Magnetic Field

The numerical solution of the equations for the total density $n$ and blueshift $\Delta E$ is given in the main text. Here we shall make a few approximations that simplify the expressions significantly and provide a physical intuition.

Firstly, we neglect processes governed by rate $\gamma_{nr}$, since their contribution is very small. We further neglect the spin relaxation processes, $\gamma_s n_{1,2}$, which at low temperatures are expected to be slower than $\gamma_r$. With these assumptions, summing the three equations at the steady state yields

$$0 = 2P - \gamma_r n_1$$

$$\Rightarrow n_1 = \frac{2P}{\gamma_r}$$

Also, from Eq. (2), we have,

$$P = R n_2 n_c$$

Plugging these values in Eq. (3), we get a quadratic equation for $n_c$:

$$\alpha n_c^2 + \beta \frac{2P}{\gamma_r} n_c + P\left(\frac{\beta}{R} - 1\right) = 0 \qquad (4)$$

The positive solution to this equation is

$$n_c = \frac{P\beta}{\alpha \gamma_r}\left[\left(1 + \frac{\alpha \gamma_r^2}{P\beta}\left[\frac{1}{\beta} - \frac{1}{R}\right]\right)^{1/2} - 1\right] \qquad (5)$$

We can take now two limits:

i. $P \ll \frac{\alpha \gamma_r^2}{\beta}\left[\frac{1}{\beta} - \frac{1}{R}\right]$: This is the limit of low power or high temperatures ($T > T_0$). We can approximate $n_c$ as

$$n_c \approx \sqrt{\frac{\beta}{\alpha}\left[\frac{1}{\beta} - \frac{1}{R}\right] P} = C_1(\alpha, \beta, R)\sqrt{P} \qquad (6)$$

We can see that at this limit $n_c$ exhibits a sublinear power dependence of $\sqrt{P}$.

ii. $P \gg \frac{\alpha \gamma_r^2}{\beta}\left[\frac{1}{\beta}-\frac{1}{R}\right]$: This is the limit of high power or low temperatures, ($T \ll T_0$). The expression for $n_c$ becomes

$$n_c \approx \left[\frac{1}{\beta}-\frac{1}{R}\right]\frac{\gamma_r}{2} = C_2(\gamma_r,\beta,R) \qquad (7)$$

We can see that at this limit $n_c$ becomes independent of power. This yields the saturated behavior of the IX energy shift with $P$.

### b. The Condensate Depletion in a Parallel Magnetic Field

The behavior of the condensate density ($n_c$) as a function of in-plane magnetic field ($B_{\parallel}$) can also be visualized from the approximate expression. It has been shown in the main text that $B_{\parallel}$ essentially shifts the dispersion in momentum space and modifies the radiative decay rate, $\gamma_r$. Hence, we can take again here two limits, of large and small $\gamma_r$.

i. At $B_{\parallel}=0$, where $\gamma_r$ is large and the power is low, we are in the limit $\gamma_r^2 \gg \frac{P\beta}{\alpha}/\left[\frac{1}{\beta}-\frac{1}{R}\right]$. This is identical to the low power limit, and $n_c$ is independent of $\gamma_r$ (see Eq. (6)).

ii. At high $B_{\parallel}$, when $\gamma_r$ becomes very small, we are in the limit $\gamma_r^2 \ll \frac{P\beta}{\alpha}/\left[\frac{1}{\beta}-\frac{1}{R}\right]$. This is identical to the high $P$ limit discussed earlier, which yields $n_c \approx \gamma_r/2\beta$. At this limit the condensate density is depleted with decreasing $\gamma_r$.

### III. Calculation Results

#### a. Behavior at Zero Magnetic Field

In order to obtain the IX blueshift from the solutions of this model, we use the expression derived in Ref. [2]:

$$\Delta E \approx 10 e^2 d^2 (n_c + n_1 + n_2)^{3/2}/\varepsilon \qquad (8)$$

where $d$ is the dipole length, and $\varepsilon$ is the dielectric constant. In constructing the solutions depicted in Figs. 3b and 4b, we used the measured values of $\gamma_r$ (Fig.2c), estimated $\gamma_{nr}^{-1} = 0.3$ ms, reported $\gamma_s^{-1} = 100$ns [3, 4], and looked for best fit for the parameters $T_0$, $\alpha_0$, $\beta$ and $R$. We find that these parameters affect different aspect of the behavior, and hence can be fitted almost independently: $T_0$ determines the temperature dependence, and $\alpha_0$ and $\beta$ – the power dependence at high and low temperatures, respectively.

Remarkably, we find approximately the same value of $T_0 \approx 7\text{K}$ for the three measured gate voltages, $V_g = -2.5\text{V}, -3.3\text{V}$, and $-3.7\text{V}$. This value is in good agreement with that obtained in the Arrhenius plot in Fig. 3a. The fitted value for $R$ is $1.2 \times 10^{-7} \text{m}^2\text{s}^{-1}$, and we took it to be temperature independent. A discussion on the other two parameters, $\alpha_0$ and $\beta$, is given in section IV.

Figure S1 shows the calculated power dependence of the steady state densities $n_1$, $n_2$ and $n_c$ at three temperatures, 4.2 K. 1.4 K and 1 K (note the different scales). The sublinear dependence of the condensate density is clearly observed at all temperatures.

### b. Behavior at a Parallel Magnetic Field

Application of a magnetic field parallel to the CQW layers, $\boldsymbol{B}_{||} = \hat{x}B$, the exciton dispersion shifts in the direction normal to it, $\hat{y}$. This shifts the dispersion minimum outside the radiative zone, resulting in an increment of the effective radiative lifetime. Fig. S2a shows the measured lifetime at $V_g = -2.5\text{V}$ as the magnetic field is increased while keeping the power and temperature fixed at $10\mu W$ and 1.5K respectively. Indeed, we find a 40 fold increase, from 40ns at $B = 0$ to $\approx 1.6\mu s$ at $B = 4\text{T}$. This should result in significant density build-up of $|\pm 1\rangle$ bright excitons, and lead to a competition between two opposite trends, namely an increase of $\Delta E$ due to the buildup of the $|\pm 1\rangle$ bright exciton density, and a corresponding decrease due to the depletion of the $|\pm 2\rangle$ dark exciton density with increasing magnetic field, results in the observed dip in $\Delta E(B)$.

Figure S2b shows the calculated steady state densities of the reservoirs (condensate and thermal bath) as a function of magnetic field. The aforementioned competition is well-manifested. The resulting blueshifts as a function of magnetic field at various temperatures are shown as solid lines in Fig. 4b of the main text. It is seen that they are in a very good agreement with the measured behavior.

In carrying out these calculations we used the same parameters for $R$ and $\alpha$ extracted from the fits to the zero field behavior. For the value of $\beta$ we used also the zero field value $\beta(0)$ and took into account the magnetic field dependence of the spin relaxation rate constant, $\beta \sim \beta(0)B^{-1/2}$ (This point is elaborated in section V). We note that the PL intensity drops with $B_{||}$ because the long lifetime of bright excitons leads to their increased diffusion to the edges and recombination there (see section VIId). Hence, we used the measured $\gamma_R(B)$ and PL intensity $I(B)$, to obtain $n_1 \sim I(B)/\gamma_r$.

### c. The Dark – Bright Exciton Mixing

A parallel magnetic field is expected to induce mixing of dark and bright excitons. Such mixing would introduce a $J_z = \pm 1$ component into the $J_z = \pm 2$ condensate, and might give rise to depletion of the condensate. The magnitude of this mixing depends on $\left(\frac{g_e \mu_B B_{||}}{\varepsilon_{bd}}\right)^2$, where $g_e$ is the electron g-factor, $\mu_B$ is the Bohr magneton, and $\varepsilon_{bd}$ is the exchange splitting [5]. Taking $g_e \approx -0.2$ (the g-factor in a 12 nm QW [6, 7]), we get $g_e \mu_B B_{||} \approx 10\mu eV/T$. The value of $\varepsilon_{bd}$ in a $10-20$ nm QW is a few tens of µeV [5], but since it depends on the electron - hole overlap, it is significantly lower in CQW, of the order of ~1µeV or less. Hence, we should obtain a complete mixing (50:50) of the bright and dark components already at very small fields, $B_{||} \ll 1T$. Consequently, if exciton mixing should yield a depletion of the condensate, it should be observed already at $B_{||} \ll 1T$. However, as can be seen in Fig. 4b we don't find any significant drop of the condensate density below 1T. This indicates that the origin of the condensate depletion is not exciton mixing.

To understand why exciton mixing does not yield a depletion of the condensate we should consider the dynamics of the system. Let us first consider the $B = 0$ case in the low temperature and low power limit, where spin flip processes are suppressed, and assume that $\varepsilon_{bd} = 0$, i.e – the bright and dark excitons are degenerate. The longer lifetime of the dark excitons should give rise to buildup of their density, and when the critical density is achieved, they condense and their density is further increased through stimulated scattering process $R n_2 n_c$. The bright excitons in this picture constitute a loss mechanism via the dark to bright spin flip processes. This implies that the dark nature of the condensate at zero field is *not* due to the very small exchange splitting, but rather due to the dynamical properties of the system: the long lifetime of the dark exciton and the suppression of spin flip processes.

The application of a parallel magnetic field doesn't change these dynamics in any significant way. Indeed, the lowest energy state of the system is not a pure dark state, but rather a superposition of $|\pm 1\rangle$ and $|\pm 2\rangle$ states. However, at steady state it decomposes into incoherent sum of dark and bright excitons, and the dynamics is similar to the zero-field case.

### IV. Discussion of the Model's Parameters

#### a. The Value of $\alpha_0$

For the calculation of $\alpha_0$ we use the derivation of the exchange integral $\xi$ of Ref. 8. From Fig. 1 of that work it follows that $|\xi|/(R_X a_X^2) = 3 \times 10^{-4}$, where $R_X$ is the exciton binding energy and $a_X$ is the exciton radius (Note that we used here the exchange integral density rather than the integral itself used in the expression of Ref. 8. Taking $R_X = 2$meV, $a_X = 10$ nm yields $\alpha_0 = \frac{|\xi|}{h} \approx 1.5 \times 10^{-8}$ m$^2$s$^{-1}$. This value is in an excellent agreement with the value obtained from the fit!

#### b. The Value of $\beta$

The spin flip mechanism, which is dominant for thermal excitons in coupled quantum wells, is the Dyakonov-Perel (DP) [9]. The main contribution to it comes from the Rashba term, which converts the crystal electric field (due to non-centro symmetry of the lattice) to an effective magnetic field. In the regular case of thermal excitons, momentum scattering with impurities changes the direction of the exciton motion in the crystal, and consequently - the direction of the effective B field changes, thereby changing the exciton precession angle. The accumulation of such scattering events gives rise to loss of spin orientation in a random walk process. The idea in our case is that for the condensate - the necessary momentum scattering is provided by collisions with the thermal excitons, and the density of impurities, which appears in the DP expression, is replaced by the density of the thermal excitons. Accordingly, the rate of this process can be expressed as $\beta(n_1 + n_2)n_c$, and should have a much weaker (non-exponential) temperature dependence. We find that $\beta \approx 2 \times 10^{-8}$m$^2$s$^{-1}$ and exhibits a weak temperature dependence, varying by ~30% over the whole range. To get an insight into the value of $\beta$, which is a scattering related spin flip rate, let us express it as $\beta \approx a^2 \gamma_s$, where $a$ is a scattering length and $\gamma_s$ is the spin flip rate. Taking $\gamma_s^{-1} \approx 100$ ns, we obtain $a \approx 40$ nm, approximately twice the exciton diameter.

### V. Measurements in Magnetic Fields

#### a. Parallel Magnetic Field

The parallel magnetic field dependence of the IX peak energy has been measured at various excitation powers, voltages and temperatures to examine the generality of the peculiar dip observed at $V_g = -2.5\ V$ and $P = 10\ \mu W$. The results are shown in Fig. S3. When the power is varied while keeping the temperature fixed at $1.5\ K$ at $V_g = -2.5\ V$, the dip is still observed, even though its magnitude decreases with lowering $P$ (Fig. S3a).

Alternatively, when the fixed power experiment (as shown in Fig. 4b in the main text) is repeated at a different voltage ($V_g = -2.9\ V$), a similar dip is observed at $T < 2.5\ K$ (Fig. S3b), where the IX blueshift due to condensate accumulation is significant (see Fig. 2a). Moreover, as the temperature increases, the dip shifts to lower $\boldsymbol{B}_{\parallel}$ values, and ceases to exist at $T = 2.9\ K$. This reflects smaller condensate accumulation at $\boldsymbol{B}_{\parallel} = 0$ with increasing $T$.

#### b. Perpendicular Magnetic Field

Figure S4 shows the shift of the IX peak energy as a magnetic field is applied along the growth direction, perpendicular to the QW plane field. The temperature, gate voltage and excitation power are fixed. Unlike the parallel field measurements, the dip is not observed, and $E_{IX}$ exhibits a quadratic diamagnetic shift. This observation rules out a significant presence of unbound electrons and holes in the system, and lead us to conclude that the $E_{IX}$ peak is due to excitons. The behavior also reconfirms the validity of the condensate-bright excitons loss mechanism, as in absence of preferential increment of bright excitons density occurring under parallel field, here we do not observe the condensate depletion and hence the dip.

### VI. Spin Relaxation of Dipolar Excitons

The relaxation time of the dipolar exciton spin in GaAs CQW is known to be much longer than the corresponding time in single quantum well (SQW) [3, 4]. The dominant process which determines the exciton spin relaxation time is the Dyakonov-Perel (DP) mechanism [9]. The DP mechanism originates from momentum scattering of the excitons. These scattering events change the precession axis of the exciton spin, around the random magnetic (and electric) fields they experience while moving in the non-centrosymmetric crystal. The discussion above applies for thermal exciton. However, for a condensate, which at mean field approximation can be considered to be at $k = 0$, momentum scattering is suppressed, and therefore, the DP mechanism is suppressed as well. Hence, one can neglect the spontaneous spin relaxation of the condensate excitons.

In the main text, we took the field dependence of $\beta$ to be $\beta(B) = \beta(0)B^{-1/2}$. Clearly, this $B^{-\frac{1}{2}}$ dependence is not valid at arbitrarily low magnetic field. As shown in Ref. 10, the validity criterion for the $B^{-\frac{1}{2}}$ behavior is $\hbar\omega_c \gg \Gamma$, where $\omega_c$ is the cyclotron frequency and $\Gamma$ is the Landau level width. In our case, we take the validity range to be $B > 1$ T, where $\hbar\omega_c \approx 2$ meV and $\Gamma \approx 1$ meV. Below this field, $\beta$ is taken to be gradually changing to the zero magnetic field value.

## VII. Nonradiative Processes

### a. Power and Temperature Dependence of Photocurrent

Fig. S5a shows the total photocurrent from the sample measured as a function of excitation power at a bath temperature of 4.2 K, at a fixed gate voltage of $-3.3$ V. We find that it is approximately linear with power, as one should expect. Remarkably, we find that for a constant excitation power, the photocurrent decreases with lowering temperature while the charge density in the wells dramatically increases (Fig. S5b). This further demonstrates that non-radiative processes do not play a significant role in determining the dark exciton density build-up.

### b. Wide Well Photocurrent Spectrum

The photocurrent spectrum in the wide well is presented at a few temperatures in Fig S6a. This is obtained using the same method as described in the main text (see the discussion corresponding to Fig. 2b, where the photocurrent spectra at the narrow well is shown). We find that the height of the trion peak and its width increase with lowering temperature. This is a manifestation of a high hole density build-up in the wide well.

In Fig. S6b we show the oscillator strength of the trion lines at both the wells, extracted by integrating the area under the trion line, as a function of temperature. We find that in both wells the oscillator strength increases dramatically between 2 K and 1 K. We further note that the trion oscillator strength at each well ceases to grow below ~ 1 K, where the IX line crosses the DX line, and the density cannot increase anymore. This provides further support to our interpretation of carrier density build-up at each well with decreasing temperature.

### c. Estimates for the Nonradiative Recombination Rate

We show in the following that the nonradiative recombination rate of the IX is $\gamma_{nr} \approx 3 \times 10^3$ s$^{-1}$, and the main contribution to it is the photocurrent.

We evaluate $\gamma_{nr}$ by measuring the blueshift at low temperature, $T = 0.5$ K and very low power $P = 10$ nW, where both spin flip terms, $\alpha n_c^2$ and $\beta n_1 n_c$, are strongly suppressed. The blueshift, $\Delta E$, is related to the condensate density by $\Delta E \approx 10 e^2 d^2 n_c^{\frac{3}{2}}/\varepsilon$ and $n_c$ can be simply written under these conditions as $n_c \approx P/\gamma_{nr}$. Hence, one can express $\gamma_{nr}$ as

$$\gamma_{nr} \approx \left(\frac{10 e^2 d^2}{\varepsilon \Delta E}\right)^{2/3} P$$

We take the absorption coefficient, $\kappa$, at the excitation energy $h\nu = 1.59$ eV as $\kappa = 1.5 \times 10^4$ $cm^{-1}$ [11], the absorption length $L = 12 + 18$ nm $= 30$ nm, and the transmission through the air-GaAs interface, $T = 75\%$. Using this to determine $P$ and the measured value of $\Delta E = 5$ meV, we get $\gamma_{nr} \approx 3 \times 10^3$ s$^{-1}$.

Let us now see how this value is related to the photocurrent. The number of e-h pairs that are generated per second at $P = 10$ μW illumination can be calculated as follows:

$$N_{eh} \approx P \times \kappa \times L \times T \times \frac{1}{h\nu} \approx 1.3 \times 10^{12} \, s^{-1}$$

The photocurrent at this power is ~100 pA. Hence, the number of electrons that are removed from the system by the photocurrent in a second is $N_{pc} \approx 6 \times 10^8 \, s^{-1}$

The ratio between the two is

$$R_1 = \frac{N_{pc}}{N_{eh}} \approx 5 \times 10^{-4}$$

The measured radiative time is ~100 ns, implying a radiative rate of $\gamma_r \approx 10^7$ s$^{-1}$. The estimated nonradiative rate is $\gamma_{nr} \approx 3 \times 10^3$ s$^{-1}$. Hence their ratio is

$$R_2 = \frac{\gamma_{nr}}{\gamma_r} \approx 3 \times 10^{-4}$$

We can see that $R_1 \approx R_2$ implying that the dominant nonradiative channel is photocurrent. We note, however, that the nonradiative losses, $\gamma_{nr} n_c$, are very small (a few percents) in comparison to the radiative losses, $\gamma_r n_1$.

### d. Spatial Diffusion to the Edges

The mesa geometry of our sample, in which the IX are not confined in the plane by a potential trap, allows them to diffuse away from the excitation regions due to the strong repulsive potential they experience. Indeed, imaging the PL signal emitted from the sample we observe a strong signal that comes from the edges of the mesa, indicating they can diffuse all the way to the edges of the mesa.

Figure S7 shows the temperature dependence of the integrated PL from the edges (blue circles) and the mesa (red circles). We find that both intensities are approximately constant in the range $2 < T < 3K$, and exhibit an opposite behavior as the temperature decreases below 2K: The edge signal sharply drops until it nearly diminishes at $T < 1$K. On the other hand, the PL intensity from the central region of the sample increases in the range $T < 2$K, and levels below 1K. We understand this temperature dependence as follows: As the temperature is decreased in the range $T < 2$K, the IX energy is blueshifted and its lifetime is shortened (Fig. 2c of the main text). The shorter lifetime reduces the number of excitons that can reach the edges, resulting in lower PL intensity.

To account for this effect in our model we weighted $\gamma_r$ by the measured PL intensity, $I$.

## VIII. Figures

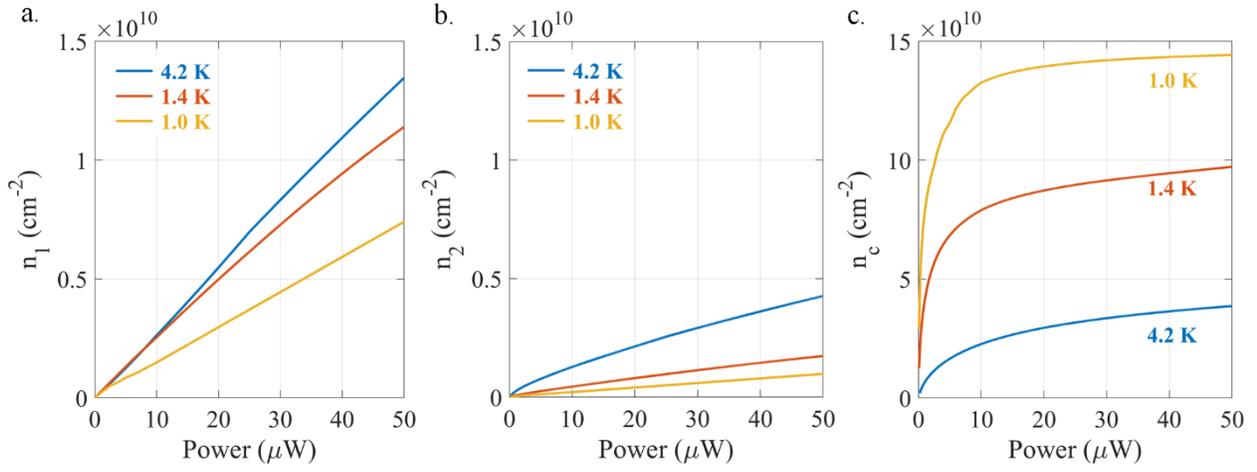

**Fig. S1.** a-c. Excitation power dependence of the calculated densities of the individual reservoirs ($n_1$ (a), $n_2$ (b) and $n_c$ (c)) at three different temperatures. The values of R, $\alpha_0$, $\beta$ and $T_0$ are taken from best fit to the data.

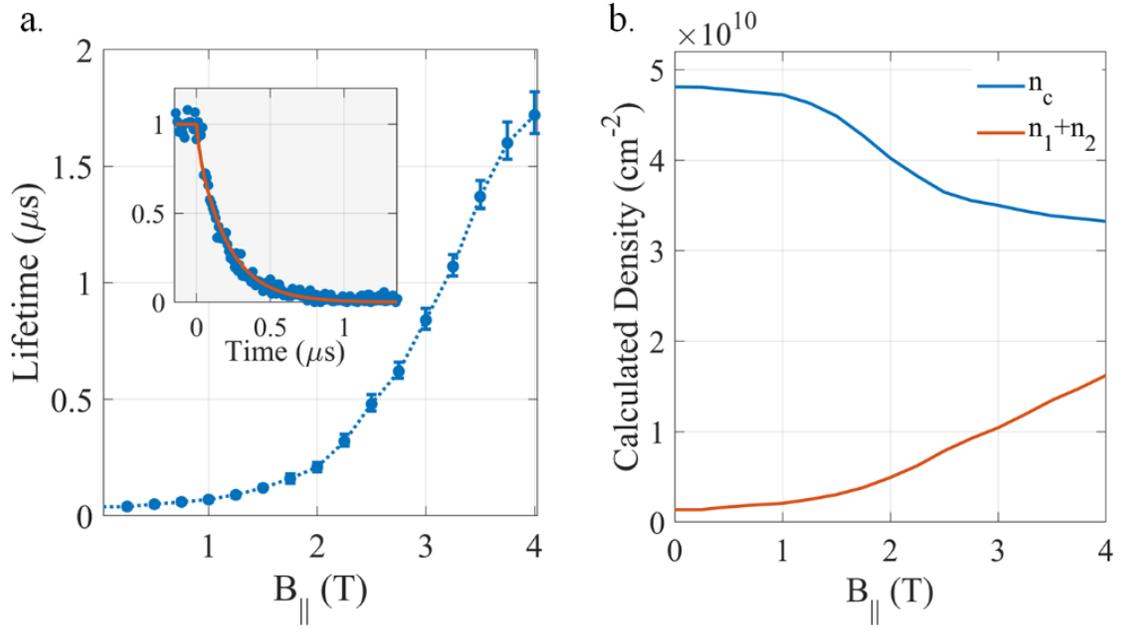

**Fig. S2.** a. Measured radiative lifetime as a function of in-plane magnetic field at $V_g = -2.5V$, $T = 1.5K$ and $P = 10\mu W$. (inset) An example of a time-resolved measurement data (filled circles) at 2T and a fit to exponential decay, with fall time 210ns (solid line); b. The calculated B-dependence of density of the condensate, $n_c$, and the thermal part, $n_1 + n_2$.

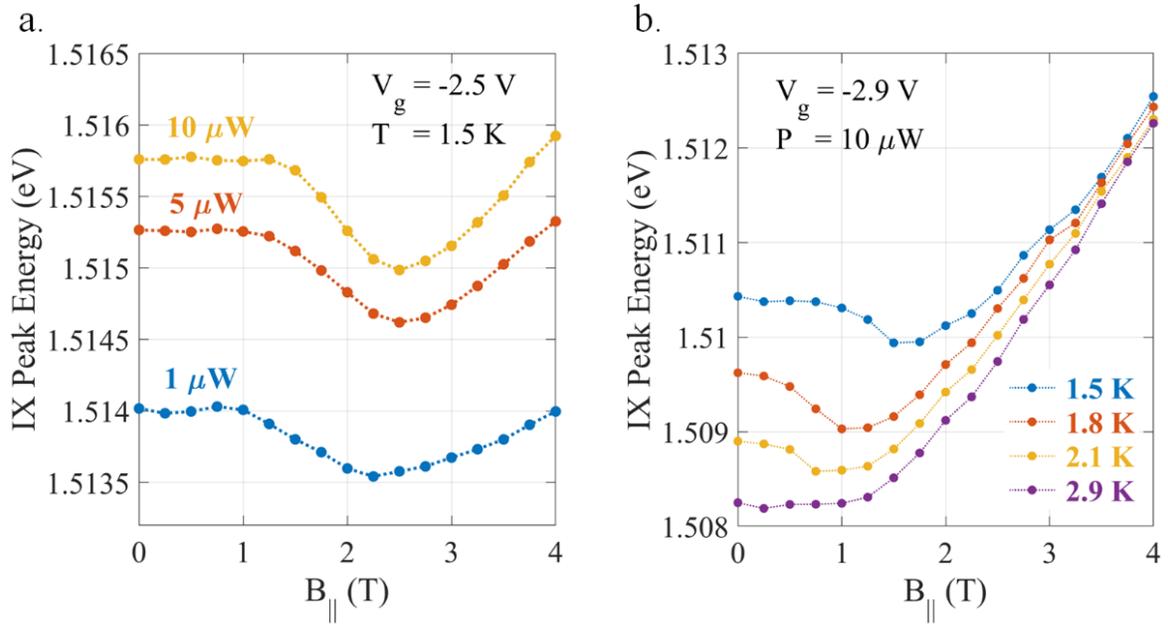

**Fig. S3.** a. IX peak energy as a function of in-plane magnetic field measured at three different powers at $V_g = -2.5$ V, T = 1.5 K; b. The same measurement at $V_g = -2.9$ V with the power fixed at 10 µW at a few temperatures.

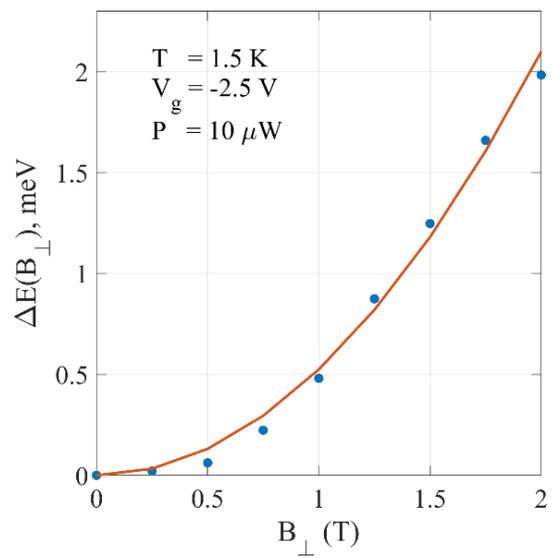

**Fig. S4.** IX blueshift relative to the zero magnetic field value, ΔE, as a function of perpendicular magnetic field. The red solid line is a quadratic fit.

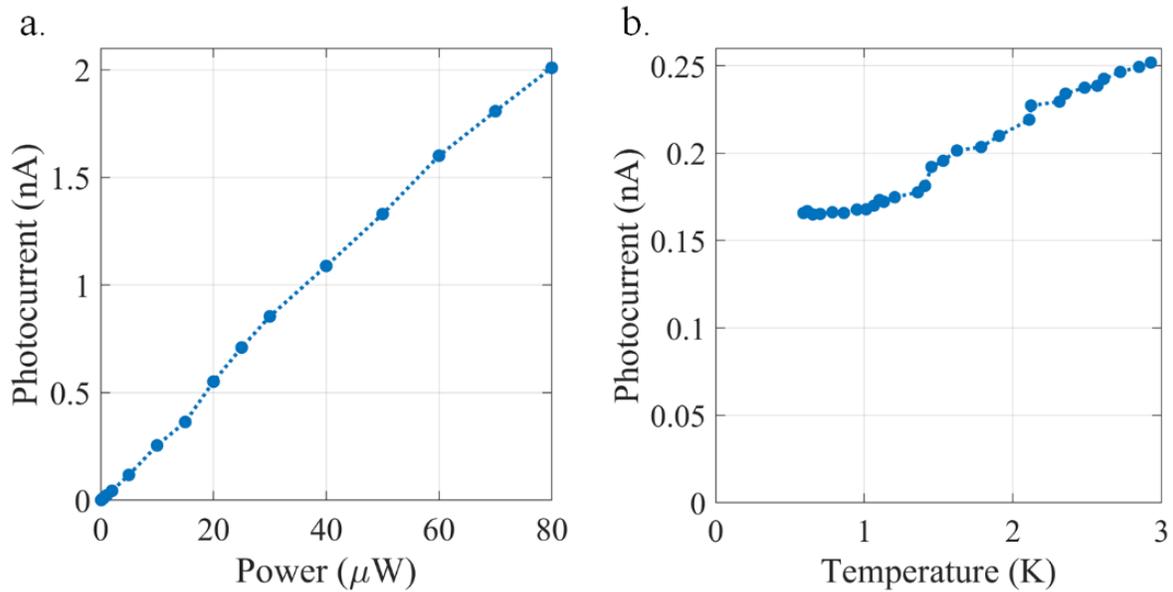

**Fig. S5.** a. The photocurrent from the sample as a function of excitation power at bath temperature, $T_{bath} = 4.2$ K. b. The photocurrent as a function of $T_{bath}$ at a fixed power of 10 µW. In both the measurements, $V_g = -3.3$ V.

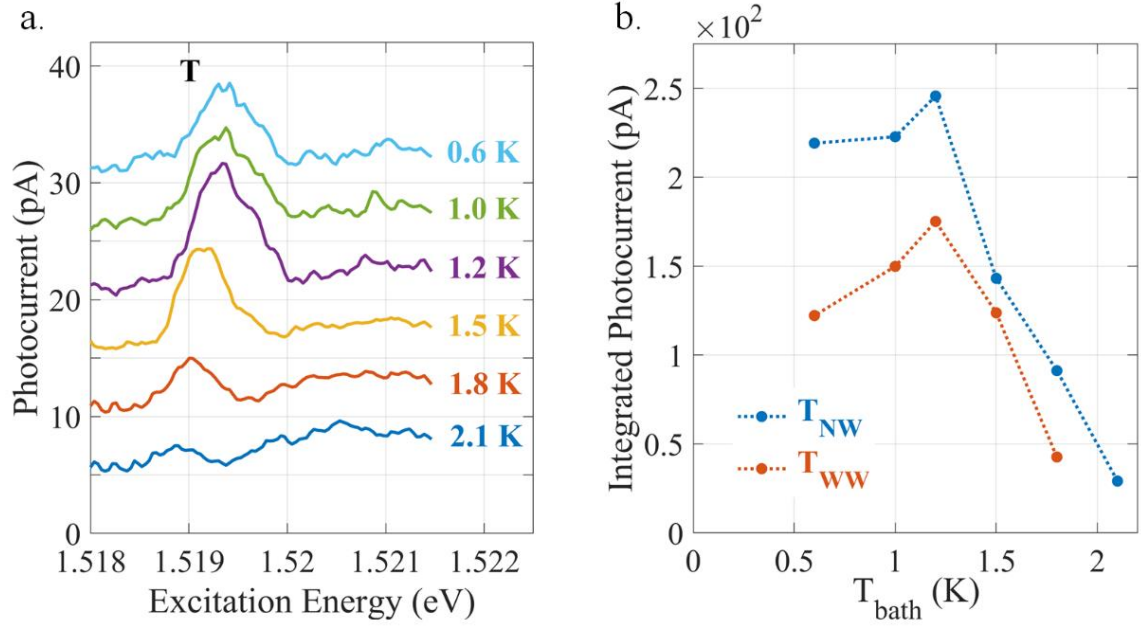

**Fig. S6.** a. Photocurrent spectra as the illuminating laser energy is tuned across the wide-well exciton resonance at various temperatures in the range 0.6 K ≤ T ≤ 2.1 K. Here $V_g = -3.3$ V and $P = 2$ μW. The spectra are vertically displaced by a fixed amount of 5 pA for clarity. b. The corresponding integrated oscillator strength of the trion line (shown both in narrow and wide well).

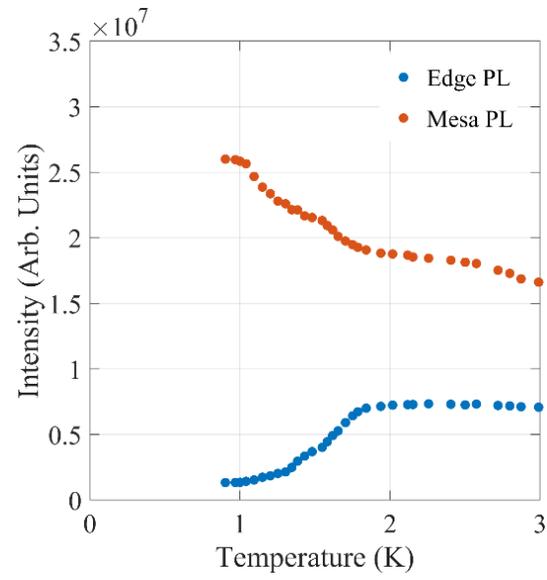

**Fig. S7.** Integrated PL intensity collected from the edges of the mesa (blue) and the central region (red) as a function of temperature.

## IX. Extended Results

Below we show representative spectra from which the data in the main text were extracted.

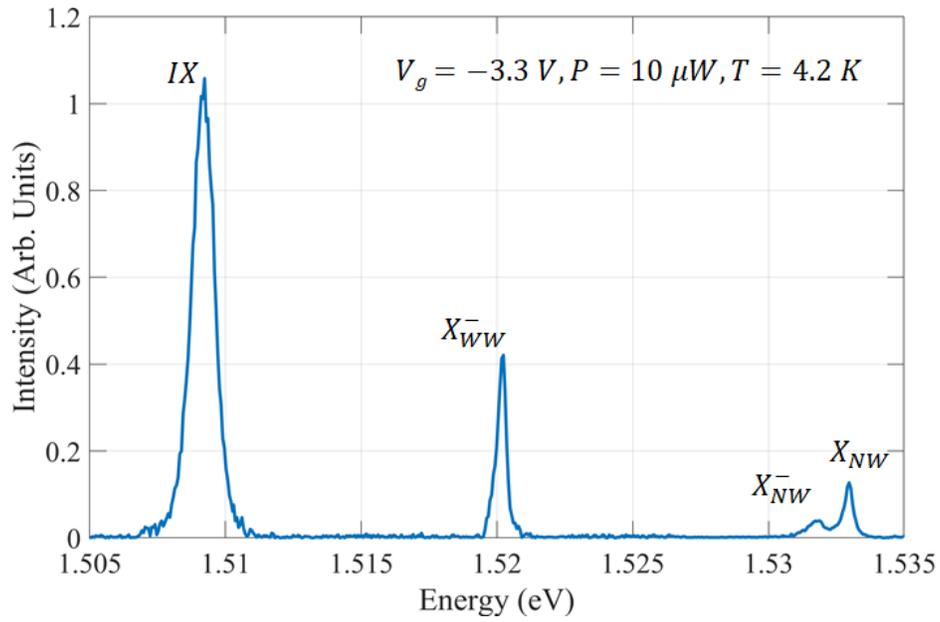

**Fig. S8.** Full PL spectrum from the mesa at a typical experimental condition. All the observable peaks are labelled.

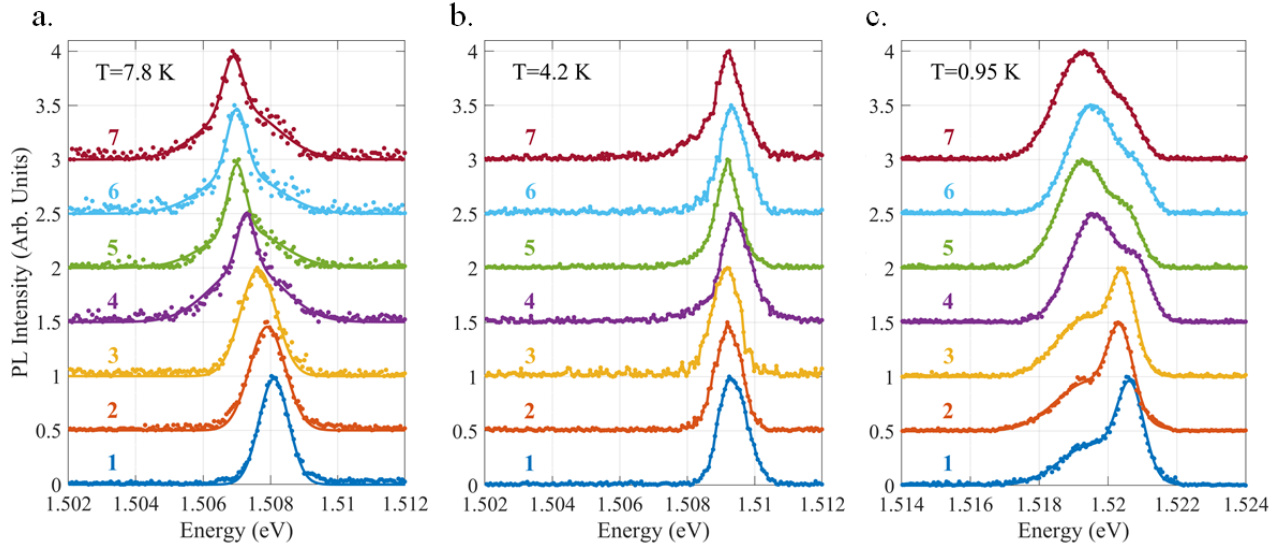

**Fig. S9**. PL spectra measured as a function of position along the mesa at three different temperatures (a. 7.8 K, b. 4.2 K and c. 0.95 K). The solid lines in a. and c. are fit to the data. The spectra are vertically shifted for clarity and normalized. The peak position of these spectra are used to obtain Fig. 2d of the main text. The numbers denote the different locations from which the PL is collected: No. 1 corresponds to the leftmost data point of Fig. 2d and No. 7 - to the rightmost data point, hence one moves away from the excitation beam while proceeding from 1 to 7. Note that the spectra at the location furthest from the center of the beam are ~ 2 orders of magnitude lower in intensity. (The energy range in c. is higher than in a and b, reflecting the large blueshift at low temperatures). In all the measurements, $P = 10 \ \mu W$ and gate voltage is kept fixed at $-3.3$ V. The broad shoulder that is observed in the 7.8 K spectra at positions 4-7, where the local excitation intensity is significantly lower than that at the beam center, is an impurity line. Indeed, this line does not shift in energy.

At the lowest temperature, $T = 0.95$ K, the IX comprises of two peaks, which exchange oscillator strength as one moves away from the excitation beam area. Nevertheless, their peak position remains unchanged. This splitting is an intrinsic property of the system, and is observed at various voltages and power values below $T \approx 1.6$ K.

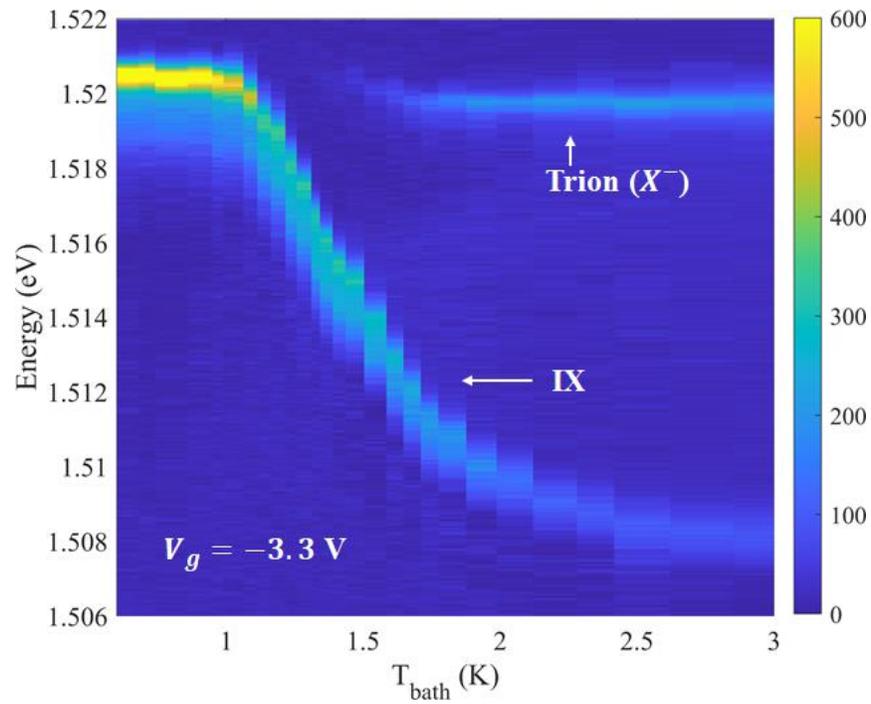

**Fig. S10**. A compilation of PL spectra collected from the whole sample as a function of bath temperature, where the excitation beam is positioned centrally on the mesa. The color code denotes the PL intensity. Here P = 10 µW and gate voltage is kept fixed at −3.3 V. The negatively charged trion ($X^-$) of the wide well and the IX are labelled.

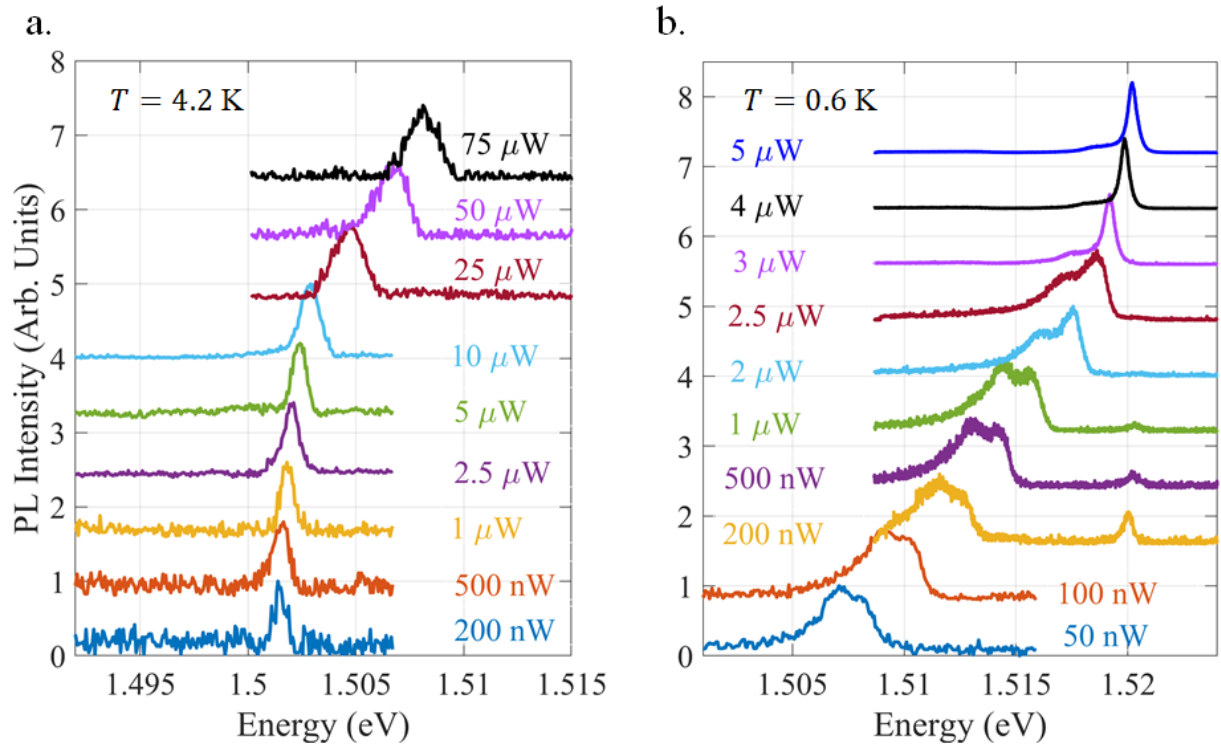

**Fig. S11.** Power dependence of the normalized PL spectra at two different temperatures: a. 4.2 K; and b. 0.6 K. In both measurements, the gate voltage is kept fixed at −3.7 V (Note the different energy scales). The sharp peak at the highest powers in b is due to the merging of the IX with the wide well exciton.

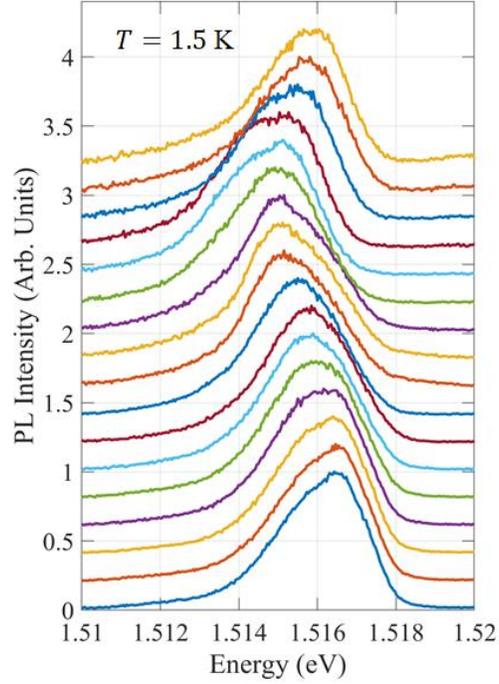

**Fig. S12.** In-plane magnetic field ($B_\parallel$) dependence of the normalized PL spectra at bath temperature of 1.5 K. Here the excitation power is 10 µW and the gate voltage is kept fixed at $-2.5$ V. The spectra are taken in the range $0\,\text{T} \leq B_\parallel \leq 4\,\text{T}$, with the spacing being 0.25 T. The bottom spectrum (blue) corresponds to $B_\parallel = 0$ T and the topmost spectrum (yellow) corresponds to $B_\parallel = 4$ T. The spectra are vertically offset for clarity.